# The protein-protein interaction network of human Sirtuin family

Ankush Sharma[a,‡], Susan Costantini[b,§], and Giovanni Colonna[a,†]

[a] Biochemistry, Biophysics and General Pathology Department and Doctorate in Computational Biology, Second University of Naples, Naples, Italy

[b] "Pascale Foundation" National Cancer Institute - Cancer Research Center, Mercogliano (AV) Italy

**Abstract**

Protein-protein interaction networks are useful for studying human diseases and to look for possible health care through a holistic approach. Networks are playing an increasing and important role in the understanding of physiological processes such as homeostasis, signaling, spatial and temporal organizations, and pathological conditions. In this article we show the complex system of interactions determined by human Sirtuins (Sirt) largely involved in many metabolic processes as well as in different diseases. The Sirtuin family consists of seven homologous Sirt-s having structurally similar cores but different terminal segments, being rather variable in length and/or intrinsically disordered. Many studies have determined their cellular location as well as biological functions although molecular mechanisms through which they act are actually little known. Therefore, the aim of this work was to define, explore and understand the Sirtuin-related human interactome. As a first step, we have integrated the experimentally determined protein-protein interactions of the Sirtuin-family as well as their first and second neighbors to a Sirtuin-related sub-interactome. Our data showed that the second-neighbor network of Sirtuins encompasses 25% of the entire human interactome, exhibits a scale-free degree distribution and interconnectedness among top degree nodes. Moreover, the Sirtuin sub interactome showed a modular structure around the core comprising mixed functions. Finally, we extracted from the Sirtuin sub-interactome subnets related to cancer, aging and post-translational modifications for information on key nodes and topological space of the subnets in Sirt family network.

‡ - ankush.sak@gmail.com
§ - susan.costantini@unina2.it
† - giovanni.colonna@unina2.it

## Author Summary

The general picture that comes out of the Sirtuins is amazing. They are hub proteins that operate deep controls in the metabolic network, and that can act in different molecular compartments where, also under the control of kinases, specific to that particular environment, recognize the correct molecular partners among the many that each of them possesses. These proteins are key nodes in different subnetwork related to cancer, aging and post-translational modifications. Therefore, the interactomic analyses on this amazing family of proteins support not only many of their known metabolic involvements in humans, which is evidence of a sound analysis, but open a window on their complex molecular mechanisms of action that, at present, is still poorly known as well as on new metabolic involvement. Thus, a systematic use of the network studies and its tools clearly open new opportunities for a better understanding of complexity.

## Introduction

Networks are emerging as valuable prototypes in simplifying complexity related to biological, social and physical sciences. Recently, some studies about the representation of complex biological systems as networks have provided key features on their structures, dynamics and functions [1-4]. In fact, the network science has developed novel paradigms including scale-free networks, small world structure and modular organization. In particular, centrality measures are useful to determine the node importance in the networks, whereas the scale-free structure indicates robustness against random failures. Modular organization is useful to separate different functions and to regulate the information transmission rate. Moreover, well-connected hubs are of high functional importance for maintaining the global network structure and communication. Some essential laws and principles governs networks which makes them useful for a deeper comprehension of biological organization, which is explained in terms of logical, informational processes and structures [5-8].

The human Sirtuin (Sirt) family is involved in many biochemical processes, as well as their pathological changes. Seven homologous proteins having structurally similar cores, which are extended by terminal segments being rather variable in length and having intrinsically disordered regions, compose this family. Sirt-1 has the highest degree of structural disorder as demonstrated recently [9]. The seven Sirts have different cellular distribution and biological functions [10]. Sirt-1 is defined as a nuclear protein involved in inflammation and neuro-degeneration processes deacetylating PGC-α (peroxisome proliferator-activated receptor gamma co-activator 1-alpha), FOXOs (forkhead box transcription factors), NFκB (nuclear factor kappa-light-chain-enhancer of activated B cells) and other nuclear substrates. Recent studies suggest the nucleo-cytoplasmic shuttling of Sirt-1 upon oxidative stress [11-12]. Sirt-2 is generally localized in the cytoplasm, and is involved in cell cycle and tumorigenesis [13]. Sirt-3 has a mitochondrial protein and a nuclear localization, but it is transferred to mitochondria during cellular stress [14]. Sirt-4 and Sirt-5 are mitochondrial proteins with different functions. Sirt-4 is an ADP-ribosyl-transferase enzyme [15], which acts on GDH (glutamate dehydrogenase) to control the insulin secretion in the mitochondrial matrix [16] and Sirt-5 is a deacetylase which activates CPS1 (carbamoyl-phosphate synthase 1) and contributes to the regulation of blood ammonia levels during prolonged fasting [17]. Sirt-6 and Sirt-7 are nuclear proteins associated with heterochromatic regions and nucleoli, respectively [18, 19]. Sirt-6 controls DNA repair, and has an ADP-ribosyl transferase activity [18], while Sirt-7 is involved in rDNA transcription acting on RNA polymerase I [19]. While there is a substantial knowledge on the biological functions of human Sirtuins, we know much less on the molecular mechanisms through which they act within the metabolic network. In fact, complex biological systems often have less experimental access so as to determine the key nodes and variables is of importance to gather information on the state of whole complex system [20] .

The aim of this work was to explore the interaction pattern and key nodes of Sirtuins and their first and second neighbors in the human protein-protein interaction network. We uncovered interaction between top degree proteins (hub proteins) of the human Sirtuin sub-interactome. Modular analysis of the Sirtuin-interactome showed a functional segregation of Sirtuin-interaction partners.

## Methods

A human protein-protein interaction map, was collected from public databases like BioGrid, HPRD, MINT, and Pathway Interaction Database, which are curated from high-throughput datasets, as well as from individual studies on interactions published in peer reviewed journals [21-24]. This dataset was manually curated and updated by Center for Bio-Medical Computing (CBMC) at University of Verona [5]. We extracted the sub-network of human protein-protein interactions containing the first and second neighbors of Sirtuins comprising 586 nodes and 243,365 interactions among them. Cytoscape software [25] was used as a visualization tool using the Kamada-Kawai algorithm.

Net analyzer [2, 26] and Centiscape [5] were used to calculate centralities of the protein-protein interaction network such as betweeness centrality (BC) [27] and closeness centrality (CC) [28]. Przulj et al. [29] and Yu et al. [30] demonstrated the importance of bottlenecks in protein-protein interaction networks and their correlation with gene essentiality. Lin et al. [31] proposed the algorithms Maximum Neighborhood Component (MNC) and Density of Maximum Neighborhood Component (DMNC) for retrieving essential, hub-like proteins from protein-protein interaction networks [31-33]. Besides these measures we utilized Maximal Clique Centrality (MCC), Edge Percolated Component (EPC), Betweenness centrality, Stress and Node degree distribution measure for exploring potentially important nodes of Sirtuins-interaction maps. Clustering pattern of network is calculated with a k cutoff value equal to 2 detects densely interconnected regions having tendency to form molecular complexes in the human Sirt network [34]. Modular analysis of protein-protein interaction networks [35] was performed using the ModuLand framework which uses the NodeLand influence function calculation algorithm with the Proportional Hill module membership assignment method [36]. Proteins specific for post-translational modifications, involved in cancer progression or senescence were extracted from various databases: Phosphositeplus [37] for proteins showing acetylation, methylation and phosphorylation, Cancer Gene Census (CGC) database [38] for proteins related to the cancer with their mutations, and HAGR [39] for proteins implicated in the senescence. Kinase-specific phosphorylation sites on Sirt family of proteins were predicted by Group-based Prediction System 2 [40]. The gene-annotation enrichment analysis were mapped to nodes (Proteins) for attaining information about biological processes, molecular function, and cellular location most pertinent to Sirt network of the nodes present in the interactome was analyzed through BiNGO plug-in [41].

## Results

**First neighbors interactions of the seven Sirtuins in the human interactome**

The interaction map including the first neighbors of the seven human Sirtuins has 228 nodes and 3769 edges. In the following parts we will show the sub-networks of this Sirt-interactome centered on each of the seven human Sirtuins.

The first neighbor interaction map of Sirt-1 included 136 nodes and 1504 edges (Figure 1) as already reported in our recent work [42]. The clustering coefficient of this network was 0.719, the mean shortest path length between any two proteins was 1.836 and the average degree was 22.11. The analysis of the putatively important proteins of Sirt-1 sub-network, detected on the basis of betweenness centrality, bottlenecks and top degree nodes, resulted in the discrimination of Sirt-1 itself and its key neighbors constituting different clusters pointing towards entirely different functions (Table S1), i.e. HSPD1 (heat shock 60 kDa protein 1), YBX1(Y box binding protein 1), HSP90AB1 (heat shock protein 90 kDa alpha, class B member 1), EEF1A1 (Eukaryotic translation elongation factor 1 alpha 1), and PRMT1 (protein arginine methyl-transferase 1). In details, Mitochondrial HSPD1 chaperonin localized outside mitochondria controls several steps in folding, transportation and assembly of proteins and may act in the innate immune system as a signaling molecule [43], HSP90AB1 is responsible for protein folding, stress induced refolding of proteins, degradation, morphological evolution, and also shows intrinsic ATPase activity [44], YBX1 over-expressed in cancer cell lines resistant to cisplatin has conserved cold shock domains and unique DNA binding domain that regulates gene expression [45], EEF1A1 plays a pivotal role in protein synthesis and delivery of all amino-acyl-tRNAs to the ribosome, and PRMT5 regulates many processes by post translational modifications such as chromatin structure, signal transduction, DNA repair, Protein translocation and transcriptional control [46].

Sirt-2 showed direct interactions with 78 proteins through 1138 connections (Figure 1). This network showed very high clustering coefficient of 0.827 with a path length of 1.6 and an average degree of 29.17. The proteins present in the Sirt-2 interactome resulted involved mostly in acetyl CoA metabolism and catabolism, in oxidative stress (keratin 1 and peroxiredoxin 2), in processes related to organ development and muscle development (Myogenic differentiation 1 and EP300), in transcriptional co-activators and peptidyl lysine acetylations (lysine acetyl-transferase 2A and 2B), in histone deacetylation, protein amino acid deacetylation and covalent chromatin modification and responses to chemical stimulus (Histone deacetylases 6).

Sirt-3 direct interactome comprised 31 nodes and 213 edges with a high clustering coefficient of 0.822 (Figure 1) and showing an involvement in different metabolic processes like ATP binding, metal ion binding and hydrolase activity. It presented densely interconnected nodes were observed in case of Sirt-2. In particular, HSPD1 (heat shock 60 kDa protein 1) resulted in the most bottlenecked node of the network being also a clique protein, which forms a cluster between Sirt-3 and Sirt-1 and involved in unfolded protein binding with other clique proteins like HSPA5 (heat shock 70 kDa protein 5), HSPD1, CMYA5 (cardiomyopathy associated 5) and HSPA1L (heat shock 70 kDa protein 1-like) and non clique proteins like HSP90B1 protein. Moreover, another important clique protein, part of a subset between Sirt-1 and Sirt-3, was FOXO3 that can be modified by posttranslational modifications and intensely associated with aging and tumor suppressor functions [47]. On the other hand, XRCC6 (X-ray repair complementing defective repair in Chinese hamster cells 6), being involved in DNA binding, resulted a clique protein among Sirt-1, Sirt-6 and Sirt-3 whereas IDE (Insulin degrading enzyme) that is the clique protein between Sirt-3 and Sirt-4 represented a possible link between ageing, diabetes, and neuro-degeneration [48-49].

Sirt-4 was interacting with all Sirtuin family as well as with the solute carrier family 25 member 5 and 6 (SLC25A5 and SLC25A6) and IDE (Figure 1). Network statistics were not calculated due to small number of interactions. SLC25A5 and SLC25A6 were clique proteins with Sirt-1, Sirt-2, Sirt-4 and IDE. IDE is a widely expressed zinc metalloprotease that regulates both cerebral amyloid β peptide levels and plasma insulin levels in vivo; in fact, it is linked to both Alzheimer disease and diabetes mellitus [50], while SLC25A6 and SLC25A5 are mitochondrial solute carrier for ADP/ATP and transport ATP in the cytosol and ADP in the mitochondrial matrix. Knowing that Sirt-4 acts as a regulator of the insulin secretion in response to glucose [51] and that the glutamate is a key neurotransmitter in brain and, furthermore, that 83% decrease in the level of free glutamate was found in subjects with Alzheimer's disease [52], it is possible to suggest that the Sirt-4 controls the glutamate levels in the brain.

Sirt-5, another mitochondrial member as Sirt-3 and Sirt-4, resulted to interact with all the members of the Sirtuin family, as well as with RELA (V-rel reticulo-endotheliosis viral oncogene homolog A) and CPS1 (Figure 1). In details, RELA interacts with Sirt-1, Sirt-5 and Sirt-6 and is associated to negative regulation of metabolism and RNA biosynthesis, and to the activation of NFKB transcription factor [53], while CPS is involved in the urea cycle intermediate metabolism, and in the arginine biosynthesis and was already indicated as substrate of Sirt5 [54].

Sirt-6 showed interactions with XRCC5, XRCC6, PRKDC (protein kinase, DNA- activated, catalytic polypeptide), CHD3 (Chromodomain helicase DNA binding protein 3), VIM (vimentin), CCNDBP1 (Cyclin-D1-binding protein 1), RELA as well as with all the Sirt family members (Figure 1) which are involved in DNA metabolism, chromosome organization and biogenesis. In particular, XRCC5, XRCC6 and PRKDC are associated with non-recombination repair [55], while Sirt-2, XRCC5, RELA, XRCC6 are responsive to stress along with some heat shock proteins [56]. Moreover, VIM, PRKDC and CHD3 are involved in intermediated filament based process and in organelle organization and biogenesis, where CCNDBP1 (cyclin-D1-binding protein 1) is found to be important for immune cell signaling [57], and CHD3 in zinc binding functionality [58].

Sirt-7 resulted to interact with HIST2H2AC and HIST2H2BE (histones cluster 2 H2AC and H2BE), UBTF (upstream binding transcription factor), POLR1A (polymerase RNA I polypeptide A), MAGED1 (melanoma antigen family D 1) and all Sirt member family (Figure 1). This Sirt-7 was responsible for the biological processes related to metabolism like macromolecule metabolism, biopolymer metabolism, proline biosynthesis and metabolism, and glutamine family amino acid metabolism. In details, UBTF was involved in transcription from RNA polymerase1 promoter [59], POLR1A in transition metal ion binding [60], MAGED1 in the p75 neurotrophin receptor mediated programmed cell death pathway [61] whereas HIST2H2AC and HIST2H2BE in the compaction of chromatin into higher order structures [62].

The compartmentalization of the direct network of Sirtuins showed the different distribution of these proteins (nucleus or cytoplasm or mitochondria or other cellular compartments) and their involvement in different functions such as DNA binding, catalytic activity, regulation of transcription activity and hydrolase activity. In details, 42.35% of proteins showed compartment specificity in nucleus or cytoplasm (Figure S1).

**Second order interactions for the seven Sirtuins**

We analyzed the second order Sirtuin interactions which includes 5786 nodes and 243365 edges (interactions) (Figure 2), as well as its topological properties, to obtain information on protein function and to understand their role and relative positions in human proteome [63]. The plot of the node degree distribution showed a decreasing trend demonstrating that Sirt family network has scale free property where bulk of peripheral nodes showed molecular function associated with transition metal ion binding and zinc ion binding (Figure 2). It also suggested occurrences of modules, i.e., subnetworks, whose members were highly interconnected but with few links to nodes outside the module. Moreover, the networks showed successive interconnected layers or inter-nested communities, with an hierarchical organization where the sparsely linked nodes were part of highly clustered areas, with the links between the different modules (named as community structures) maintained by few hubs [64]. The clustering coefficient graph showed a decreasing trend and the value related to the network heterogeneity, which accounts for the variance of the connectivity, reflects the tendency of a network to contain hub nodes (Figure 2) [65].

Sirt network showed an increasing trend of the neighborhood connectivity distribution, which reports the average of the neighborhood connectivity of all proteins (*n)* with *k* neighbors (Figure 2). The related slope, equal to 0.405, evidenced the presence of high degree nodes, known as hubs. Moreover, we evaluated also the related assortativity coefficient (r), which ranges between -1 and +1 and is related to the preference for a network's nodes to others that are similar. In Sirt network, this coefficient resulted equal to 0.619 indicating that our network showed assortativity with a correlation between nodes of similar degree.

Moreover, we evaluated the betweenness centrality that provides inferences on the importance of proteins on the basis of load placed on the given node in the network, and, hence, information about the core skeleton of the network. Betweenness centrality demonstrated increasing trend with maximum load placed on: i) TP53 (tumor protein p53), which is DNA binding tumor suppressor protein, ii) UBA52 (ubiquitin A-52 residue ribosomal protein fusion product 1), which is involved in the maintenance of chromatin structure, the regulation of gene expression, and the stress response, and iii) EEF1A1 (eukaryotic translation elongation factor 1 alpha 1), which is a protein responsible for the enzymatic delivery of aminoacyl tRNAs to the ribosome (Figure 2). Then, we focused our attention on nodes showing hub–hub interactions and we calculated whether these nodes exhibit rich club property [66]. Twenty-five proteins exhibited hub-hub interactions in network despite rich club coefficient less than one. Core nodes of central module (EEF1A1 and UBA52) shape the core skeleton of network (high betweenness central proteins) with large number of short path lengths crossing through these proteins allows us to infer about the faster information transfer at the core and the rigidity associated with the networks (Figure S2). The resilience of network skeleton was examined concerning i) 50 core community centrality proteins (CC), ii) 50 betweenness centrality (BC) and 25 interlinked hub subnet were tested by deleting the proteins present in any of the two network from one of the top 50 protein subnetwork. Targeted deletion of all the proteins, which were part of the other two top-node sub networks, showed large disruption of interlinked hub subnet (deleted BC and CC) whereas other two subnets related to BC (deleted Interlinked hub subnet and CC) and CC (deleted BC and interlinked hub subnets) remained largely unaffected suggesting that hub proteins either had high information transfer proteins or were high betweenness centrality proteins which forms the core skeleton of the network (Figure 3).

**Modularization of network**

The most part of the functional activity inside cell is organized as a network of interacting modules where genes and proteins co-operatively respond to different conditions [67]. Therefore, the modular overlaps exhibit the functional diversity of proteins. We calculated by ModuLand framework [36] the community centrality values corresponding to proteins showing the influence of the Sirts interactome on the given protein, and, hence, the level of importance of the protein in the whole Sirt interactome. The nodes associated with high community centrality on the selected level (whole Sirt network) form the core of module of the interactome. 20 overlapping modules of the sirtuin-network were detected using the ModuLand plug-in for Cytoscape [36].The modular structure was organized around a core with mixed functions (Table 1) where EEF1A1 module had the highest modular assignment value which indicates many cores having interlinked hub nodes (Table S2). Moreover, CSNK1A1 (casein kinase 1 alpha 1), HDAC1 and NDUFA10 (NADH dehydrogenase ubiquinone 1 alpha subcomplex 10) involved in cytoskeletal signaling (including microtubule reorganization), transcriptional regulation (including chromatin remodeling) and mitochondrial terminal oxidation and ATP synthesis, respectively. Other detected overlapping modules were detected to play a major role in the integration of cellular responses, in the protein and RNA binding, in the signal transduction, and in the enzyme regulation and transferase activity.

**Sirt family network involved in Cancer subnet**
Sirtuins are widely known as critical regulators of aging and cancer and Sirt-1 was suggested to act as double sword in cancer [68], furthermore, we focused our studies also on oncogenic mutations, which were reported in Cancer Gene Census (CGC) database [38]. Subnet related to cancer in SIRT network was composed of 302 proteins out of 468 proteins present in CGC (Figure 4). The centrality statistics and modularization of network were calculated only for the connected components containing 279 nodes and 1677 interactions. The clustering coefficient was comparatively less than that obtained in SIRT first order and SIRT second order interactome. In fact, the value obtained for cancer related proteins was equal to 0.371 with average number of neighbors of 12.10 and path length of 2.70 (Figure S3). We calculated the top ten proteins based on the different algorithm with high centrality statistics listed in Table 2A, where we find five hub proteins directly interacting with SIRT family, i.e. EP300, JUN, RB1 (retinoblastoma 1), TP53 and EWSR1 (Ewing sarcoma breakpoint region 1). The analysis of the modularization evidenced that Sirt network involved in cancer has three overlapping modules with the following core nodes: EP300, ERCC6 (excision repair cross-complementing rodent repair deficiency complementation group 6) and XPA (xeroderma pigmentosum complementation group A). EP300 exhibits transcriptional regulation activity, ERCC3 and XPA showed high inter modular links, and, hence, exhibit overlaps with similar functionality of damaged DNA binding. The other low degree vertices of the cancer sub-network were sparsely distributed contributing to the high average path length and to lower values for centrality indices (like for example the clustering coefficient) (Table 3). Cancer proteins were more represented among housekeeping genes [69]. In particular, PPP2R1A (protein phosphatase 2 regulatory subunit A alpha) and MYC (v-myc myelocytomatosis viral oncogene homolog) were high information transfer proteins in cancer subnet of Sirt family network showing high community centrality in top 5% of proteins in whole SIRT family network.

**Aging subnet in Sirt interactome**
We analyzed the proteins related to aging from human genomic aging resource dataset [39] in Sirt family interactome. The obtained network comprised of 198 proteins and 2506 interactions (Figure 5). All the statistical analysis related to node degree distribution, clustering coefficient and topological coefficient showed a decreasing trend with, in particular, a high clustering coefficient of 0.454 and the average number of neighbors equal to 25.303 (Figure S4). Centrality statistics of aging network in SIRT interactome evidenced that i) TP53 showed largest value for betweenness centrality that illustrates the large load placed on the node as larger number of short path traverse through node and ii) TP53 occupied central positions within the communities to which it belong. Another central node was YWHAZ (tyrosine 3- mono-oxygenase/tryptophan 5-monooxygenase activation protein zeta polypeptide) that formed the core skeleton of aging network (Table 2B). Modular overlapping was not significantly observed rather short average path length between the nodes and larger homogeneity on combined basis suggested aging sub network easily synchronizable [70-71]. Since age progression is accompanied with the chronic inflammatory diseases leading also to cancer, we have hypothesized that the same proteins can be deregulated or closely associated with the perturbed proteins in the network. In fact, in SIRT second order network, forty-two proteins were implicated both in aging as well as cancerous conditions, and 61% of these proteins exhibited DNA binding activity. Moreover, MYC, TP53, WRN (Werner syndrome, RecQ helicase-like), RB1, EP300 and JUN had experimentally evidenced interactions with Sirts.

**Kinase subnet**
Phosphorylation and dephosphorylation are essential for eukaryotic signaling and about 30% of proteins was phosphorylated and dephosphorylated at a given time governing many physiological processes, which upon deregulation can cause cancer and other diseases [72]. We extracted the sub-networks of 117 kinases linked with 1366 edges present in Sirt family interactome (Figure 6). Moreover, the network resulted to have faster information flow with least average shortest path length of 1.9 analyzed in all subnets and average number of neighbors of 25.06 (Figure S5). The average clustering coefficient for the network was equal to 0.664 and distribution showed negative slope where as node degree distribution, showed a decreasing trend stating kinase subnet to be robust (Figure S5). Compartmentalization of kinase subnet inferred about presence of kinases at multiple location, i.e. 85 in cytoplasm, 63 in nucleus and 22 in cytoskeleton. Gene ontological analysis highlighted that 82% of the proteins in kinase network were implicated in regulation of cellular processes and 38% out of 117 kinases present in second order SIRT interactome was involved in response to stress. In details, cyclin-dependent kinase 1, 3, 4 and 6 (CDK1, CDK3, CDK4 and CDK6), PLK1 (polo-like kinase 1), and ATM (ataxia telangiectasia mutated) were involved in mitotic cell cycle. Consequently, we extracted the proteins present in first order interactome evidencing that the network contained a bulk of central proteins that were the potential substrates of 117 kinases. Insights on the kind of interactions possessed by kinome in Sirt interactome were dealt through the gene ontological data and centrality statistics. The Kinome interacts with proteins like TP53, RELA, JUN and MEF2D (myocyte enhancer factor 2D). CDK1 and GS3K (glycogen synthase kinase 3) were directly interacting kinases in the Sirt interactome. ATM, EGFR (epidermal growth factor receptor), FGFR1 (fibroblast growth factor receptor 1), JAK 2 (janus kinase 2) and TTK were the kinases having post-translational modifications and were implicated both in ageing and cancerous condition. In Sirt interactome, the kinases exhibited acetylation property and implicated in cancer and aging subnets; in fact, FGFR, JAK2 and EGFR kinase showed acetylation. The acetylation of kinases suggests only two possible interaction ways, one where Sirts deacetylase kinases and a second one where kinases phosphorylate Sirts. Experimental identification of phosphorylation sites is labor-intensive and often limited by the availability and optimization of enzymatic reaction. Computational methods of prediction may facilitate the identification of potential phosphorylation sites; thus, to investigate the importance of the phosphorylation in all the Sirtuins, we extrapolated this information from their second order network. In particular, Sirt1 interacts with 106 kinases, Sirt2 with 95, Sirt3 with 68, Sirt4 and Sirt5 with 17, Sirt6 with 74 and Sirt7 with 22 (see also Table S3 for details).

**Acetylation and methylation subnets**

We extracted in Sirt interactome the subnetwork of proteins having acetylation property. It comprised of 1367 proteins and 44873 interactions and showed scale free property with average number of neighbors of 65.65. Hub proteins like EP300, RELA, and POLR2A were located at the centre of this network, together with two kinases, ATM and TTK, that possessed methylation property, and EGFR that showed acetylation. However all these proteins were implicated also in aging as well as cancer subnetworks. Remarkably, 71 of these proteins resulted to be involved also in methylation and, between these there were also important housekeeping hub proteins like TP53, EP300, NCOA2 (nuclear receptor coactivator 2) and DNMT1 (DNA cytosine-5-methyltransferase 1) (Figure S6). In particular, the hub protein histone acyl transferase EP300 resulted the highest modular bridgeness, showing both acetylation and methylation property and being implicated also in cancer and aging subnets. This protein interacts directly with Sirt-1 and was involved in biological processes related with chromatin silencing at telomere. Other acetylated and methylated proteins with high centrality properties resulted NCOA, MYC and CLOCK (circadian locomotor output cycles kaput) that with the Sirts potentially linked epigenetic regulation

**Conclusions**

Biological processes inside our body are governed by well-defined organization of proteins into complexes, which perform different functionality acting as molecular machines. The holistic vision, centered on network studies for the characterization of human diseases, redefines the field of medicine by finding new and personalized treatments different from the traditional approach which relies on simple clinical observations [73]. Protein-protein interaction (PPI) networks include small interwoven networks inside them. These small interwoven networks contain functional information on complex biological network and interaction between proteins comprises information related to biological processes of the interactants.

In our studies, we focused on the molecular interaction maps of the important protein family of the human Sirtuins, which is involved in many important molecular functions and biological processes. The analysis of the first order interactions for all seven Sirts evidenced that, i) the Sirt-1 and Sirt-2 maps presented a very high number of nodes and edges supporting the many experimental studies regarding these two proteins and their involvement in many important biological processes (see figures 1A and 1B), ii) EP300, essential in the processes of cellular proliferation and differentiation, was detected as hub protein, both in Sirt-1 and Sirt-2 network, iii) IDE (Insulin degrading enzyme), clique protein between Sirt-3 and Sirt-4 suggested an involvement of these two proteins in ageing, diabetes, and neurodegenerative diseases, and iv) , RELA interacts with Sirt-1, Sirt-5 and Sirt-6 evidencing that it is associated to the activation of NFKB transcription factor.

The Sirt protein interaction network resulted to cover approximately 25% of human proteome in second-degree network and exhibited scale-free and preferential hub-hub inter-connected proteins. In this network, 20 overlapping modules suggested pleiotropic functions and the top 10 core proteins showed hub-hub interactions in EEF1A1module with mixed functions such as ATP-binding, cytoskeletal organization and transcriptional regulation. In fact, the Sirt network showed a modular structure on the core which comprised mixed functions with three interrelated network structures: i) hub-hub interlinked proteins were found to be involved in important functions constituting the core module of the network and involving a large number of shortest path length and hence can contribute unevenly towards global communication in Sirt network, ii) community structure involved in processes related to binding and top community centrality proteins showed localized in multiple cellular compartments and iii) the most part of Sirt network followed peculiar hub and low vertices functional organization for providing robustness against random deleterious mutations, which might influence the metal ion binding molecular function associated with peripheral nodes. Moreover, we expanded our study to proteins involved in cancer due to somatic mutations; this sub-network showed low degree vertices with high average path length and hence inefficiency in information transfers. On the other hand, the aging sub-network showed high level of synchronizability as evidenced from the shortest average path length typical of a homogeneous network. This is strictly correlated to specific functions or dysfunctions of biological systems; in fact, in epilepsy and Parkinson's diseases, seizure activity and tremor are due to excessive synchronization [74-75] and chronic disruption in circadian system promotes aging and is prone to various disease states including cancers, heart disease, ulcers, and diabetes [76-77]. However, when we mapped the proteins common in aging as well as cancer in the second order SIRT interactome, we evidenced that 42 proteins associated with cancer and aging and six proteins (MYC, TP53, WRN, RB1, EP300 and JUN) showed direct interaction with the SIRT family interactome. This confirms that Sirt family is involved contemporaneously in chronic inflammatory processes leading both to ageing diseases and cancer. The Sirt network consists also of acetylated substrates such as transcription factors and proteins responsible for circadian rhythms, which can influence metabolic pathways and whole cellular milieu whereas aging subnet showed proteins with wide variety of posttranslational modifications ranging from methylation, acetylation and phosphorylation.

Finally, since many physiological processes and some diseases are associated with the abnormal phosphorylation and about 30% of proteins is phosphorylated and dephosphorylated in highly dynamic interactions, we analyzed the kind of biological processes and kinases associated with Sirt network. In particular, in the second order network of all Sirtuins we evidenced that Sirt-1 interacts with 106 kinases, Sirt-2 with 95, Sirt-3 with 68, Sirt-4 and Sirt-5 with 17, Sirt-6 with 74 and Sirt-7 with 22. This surprisingly high number of kinases is very interesting because the presence/absence of phosphate groups seems important in modulating the recognition of the different proteins and to regulate the enzymatic activity [78]. In a recent paper we show that most of sirtuins possess numerous phospho-sites on terminal segments [79]. This special condition should be considered taking into account that these segments are intrinsically disordered and therefore very flexible. All this leads us to consider that they represent structural regions highly exposed and available to the recognition of molecular partners; moreover, they have charged stretches in which phospho-sites are often allocated by generating pospho-isomers important for the one-to-one recognition among the numerous molecular partners that each of these proteins possesses. The general picture that comes out of the sirtuins is amazing. They are Hub proteins that operate deep controls in the metabolic network, and that can act in different molecular compartments where, under the control of kinases, specific to that particular environment, recognize the correct molecular partners among the many that each of them possesses. This mechanism of action, previously never clearly focused because of the strong focus on the study of their physiological and pharmacological effects that has practically neglected the study of the molecular basis of their action, should be extended because it is lawful to imagine that other types of post-translational changes may come into play.

In conclusion, we report also the overall statistics found in the Sirt interactome and in its different subnets (Table 3). In fact, the cancer sub-network had the highest average path length indicating the presence of low degree vertices sparsely distributed whereas its values for centrality indices, like clustering coefficient, were lesser than other subnets corresponding to aging and post translational modification. Moreover, Sirt family network showed the highest value (equal to 1.358) for network heterogeneity due to the considerable bigger size of this network and to the presence of many hub nodes. On the other hand, Cancer subnet had highest heterogeneity inferring of its larger tendency to have hub proteins. Finally, the subnets related to aging and posttranslational modifications, in particular the subnet of kinases, demonstrated smaller values for the average path length, and this means faster rate of information flow. Our Interactomic analyses on this amazing family of proteins support not only many of their known metabolic involvements in humans, which is evidence of a sound analysis, but open a window on their complex molecular mechanisms of action that, at present, is still poorly known as well as on new metabolic involvement. Thus, a systematic use of the network studies and its tools clearly open new opportunities for a better understanding of complexity [80].

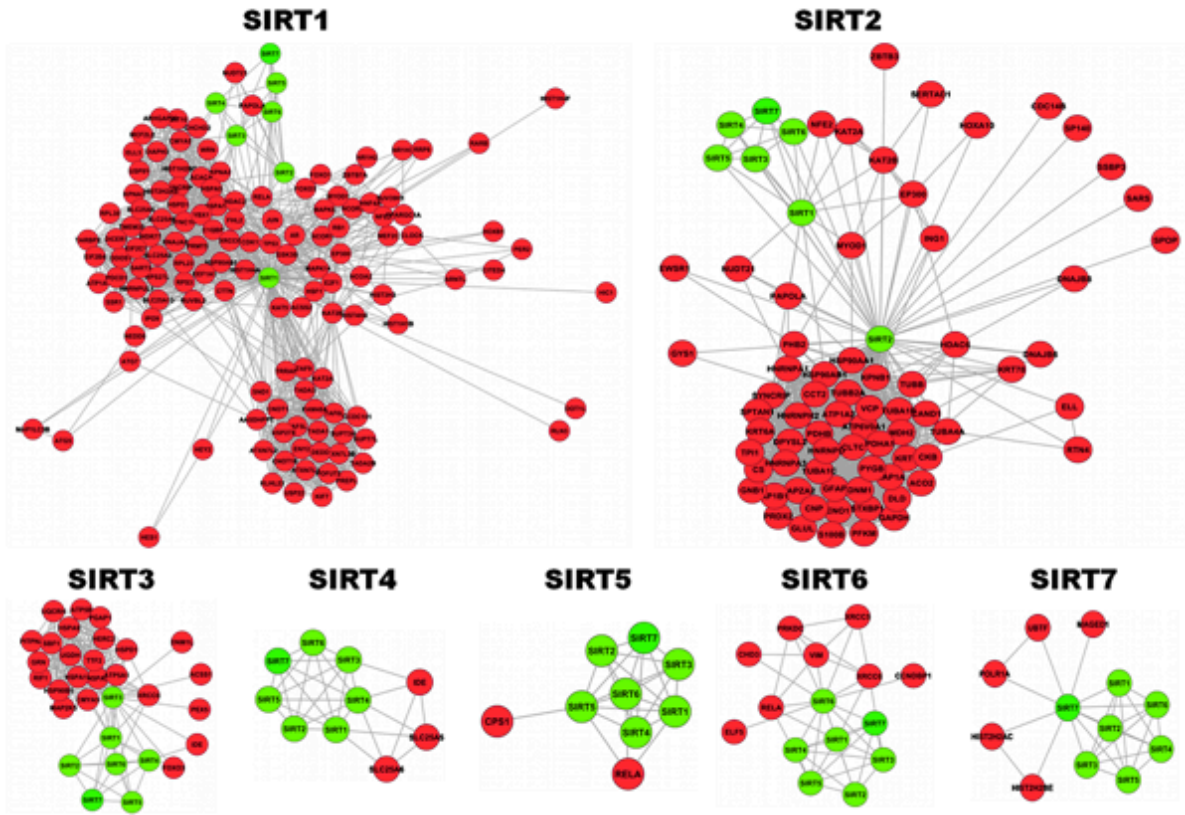

**Figure 1.** First order interaction map of seven Sirtuins where Sirts are evidenced in green and other proteins in red.

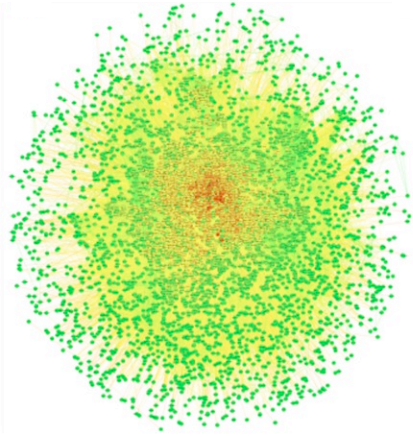
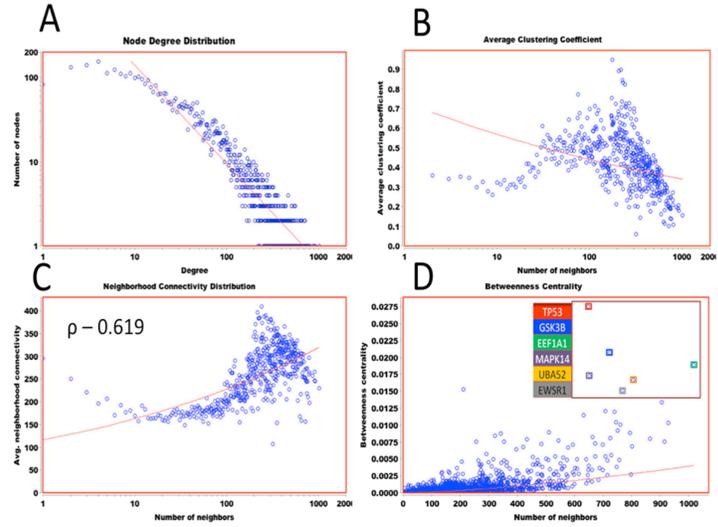

Figure 2. Sirt 2nd order interactome visualized by spring embedded layout with color coded from red → yellow → green. Red nodes illustrated proteins with larger number of neighbors and green nodes with lower vertices. Evaluation of topological properties of second order Sirtuin interactome: (A) node degree distribution (B) average clustering coefficient, (C) neighborhood connectivity distribution, and (D) betweeness centrality measure.

Top 50 betweenness centrality nodes
after deletion➔

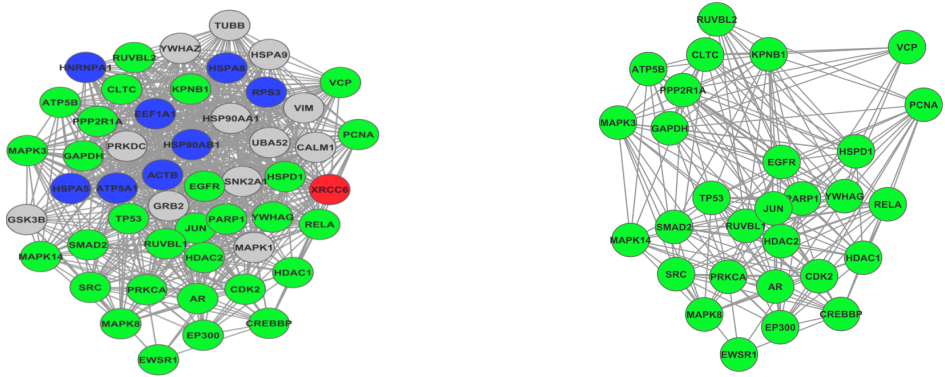

Top 50 community centrality nodes
after deletion➔

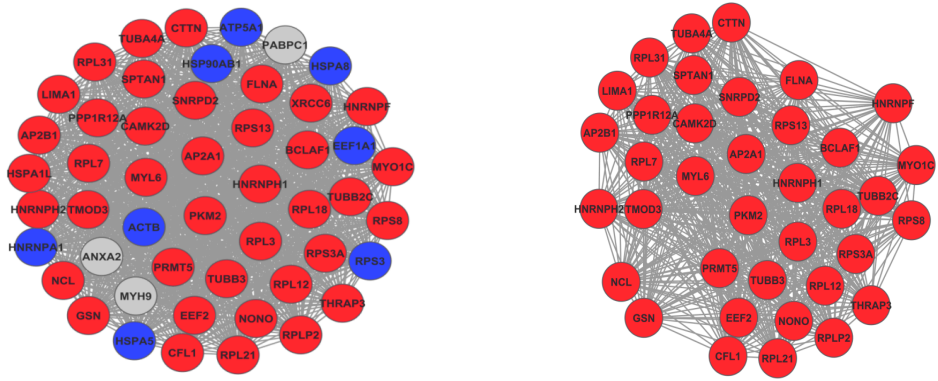

Top degree (hub) interlinked proteins
after deletion ➔

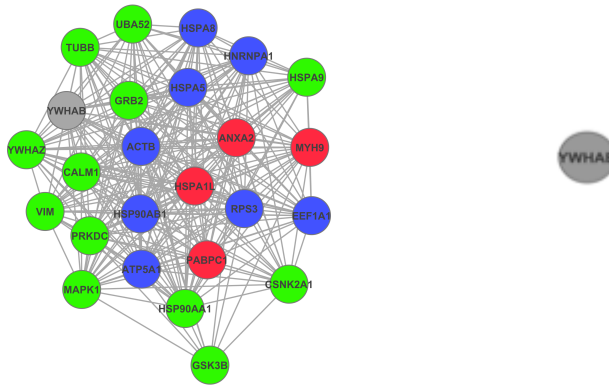

**Figure 3.** Three networks related to 50 core community centrality proteins (CC), 50 betweenness centrality (BC) and 25 high degree interlinked hubs (IH) were reported after deletion of nodes present in other two networks. i.e. subnets of community centrality and interlinked hub nodes. The community centrality node

**Cancer Subnet**

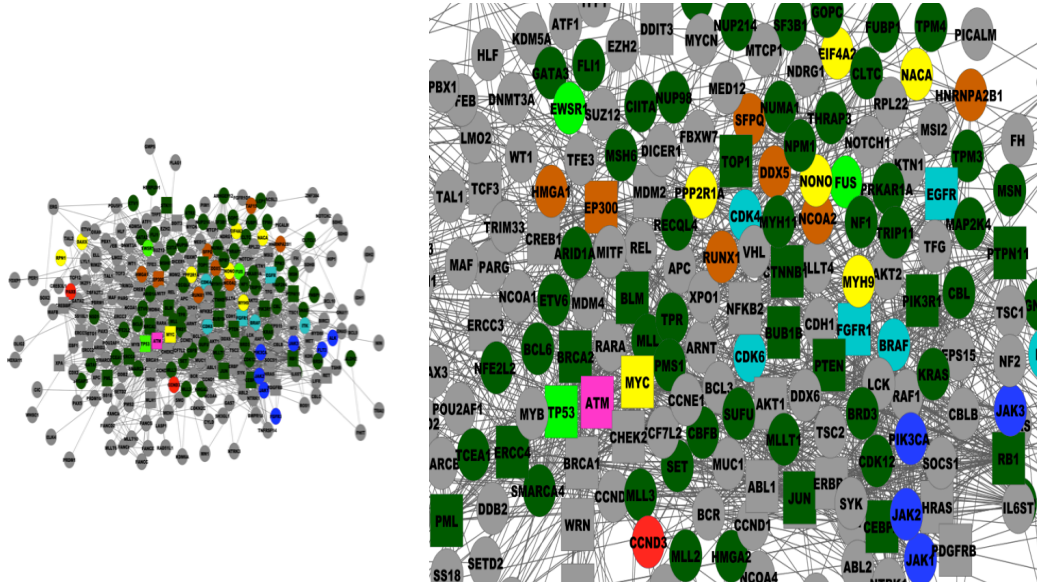

**Figure 4.** Sirt family network involved in cancer subnet in green the proteins acetylated, in blue the kinases, in fluorescent green the proteins methylated, in red the housekeeping, in orange the proteins involved in methylation and acetylation, in cyan the proteins involved in acetylation and kinases, in yellow the proteins acetylated and housekeeping, and in magenta the proteins methylated and kinases.

# Aging Subnet

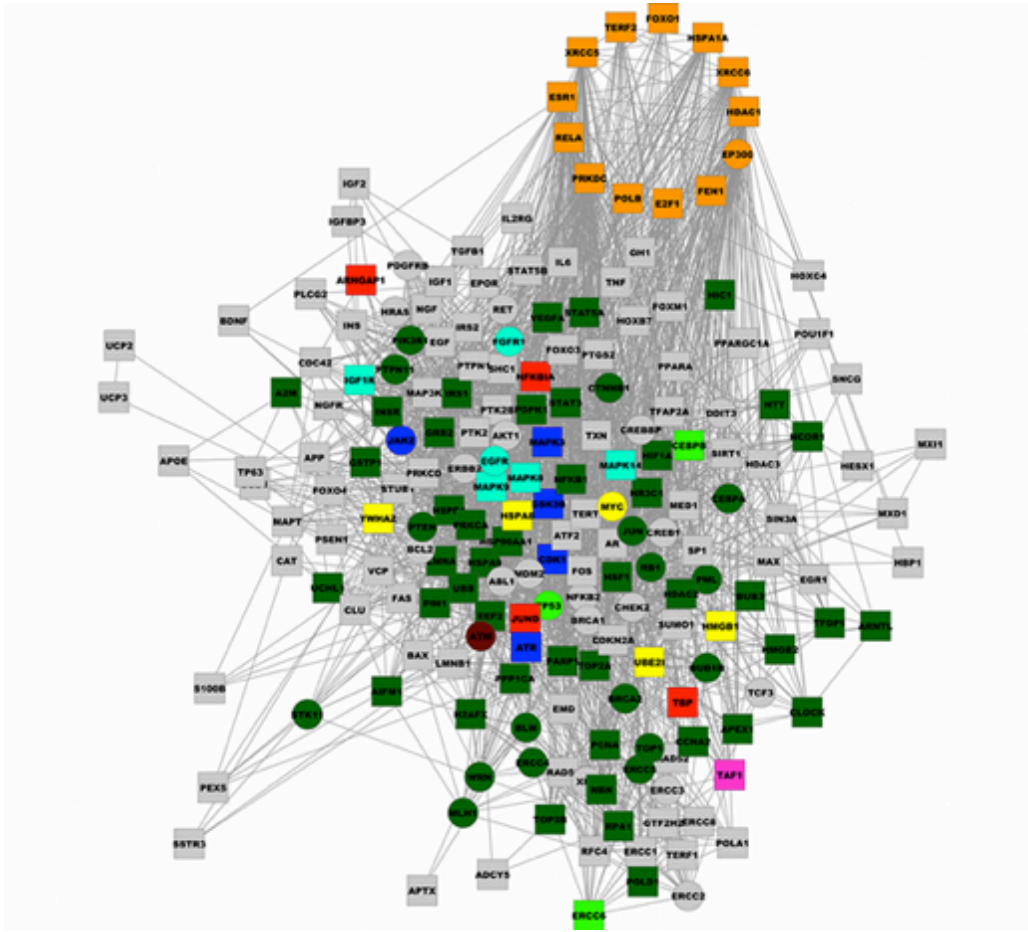

**Figure 5.** Sirt family network involved in aging subnet with in green the proteins acetylated, in blue the kinases, in fluorescent green the proteins methylated, in red the housekeeping, in orange the proteins involved in methylation and acetylation, in cyan the proteins involved in acetylation and kinases, in yellow the proteins acetylated and housekeeping, and in magenta the proteins methylated and kinases.

## Kinase Subnet

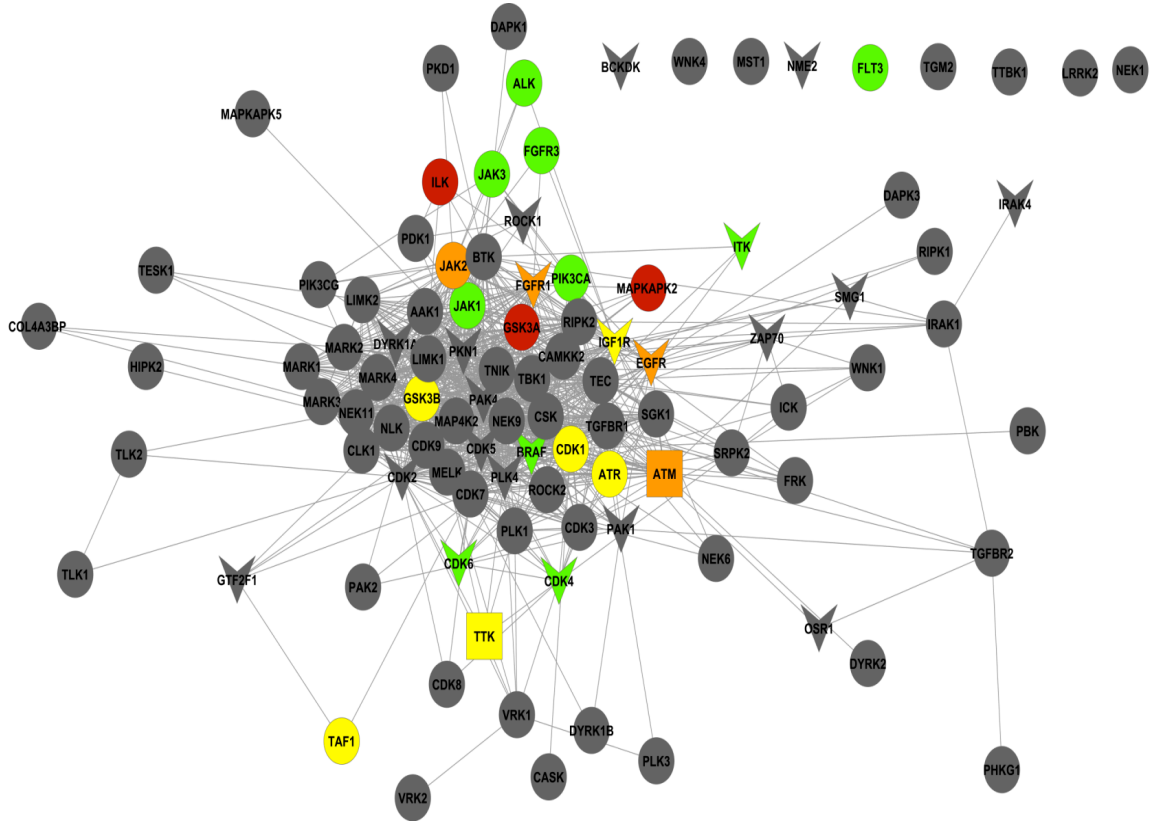

**Figure 6.** Sirt family network involved in kinase subnet with the proteins involved in the methylation by squares, those involved in methylation and acetylation by diamonds, those involved in acetylation by inverted arrowheads. Different colors indicated proteins involved in cancer (green), in aging (yellow), in cancer and aging (orange) and in housekeeping (red).

Table 1. Molecular consensus function based on GO terms associated with 20 overlapping modules detected by ModuLand framework

| Modules | Module Name | Molecular function |
|---|---|---|
| **Module 1** | **EEF1A1** | Nucleotide binding |
| **Module 2** | **CSNK1A1** | Transferase activity, transferring phosphorus-containing groups |
| **Module 3** | **HDAC1** | Transcription regulator activity |
| **Module 4** | **KRT33B** | Structural molecule activity |
| **Module 5** | **LSM2** | RNA binding |
| **Module 6** | **KALRN** | Enzyme regulator activity |
| **Module 7** | **TIAM1** | Enzyme regulator activity |
| **Module 8** | **NDUFA10** | Oxidoreductase activity |
| **Module 9** | **TERF2IP** | DNA binding |
| **Module 10** | **COPS6** | No annotation retrieved |
| **Module 11** | **MRPS21** | No annotation retrieved |
| **Module 12** | **FAM175B** | Ubiquitin binding |
| **Module 13** | **POLA2** | Nucleotidyltransferase activity |
| **Module 14** | **TUBGCP3** | Structural constituent of cytoskeleton |
| **Module 15** | **ETFB** | Nucleotide binding |
| **Module 16** | **RBPMS** | Transcription regulator activity |
| **Module 17** | **GABRA1** | Signal transducer activity |
| **Module 18** | **GABRB2** | Transmembrane transporter activity |
| **Module 19** | **SARS** | Catalytic activity |
| **Module 20** | **CNDP1** | Transferase activity |

**Table 2.** Statistical analysis on Cancer (A) and Aging (B) subnets extracted from second order Sirt interactome in terms of Degree, betweenness, stress, MCC (Maximal Clique Centrality), DMNC (Density of Maximum Neighborhood Component), MNC (Maximum Neighborhood Component), EPC (Edge Percolated Component). The proteins resulted hub are evidenced in bold and underlined.

**A)**

| Sr. no | Degree | Betweenness Centrality | Stress | MCC | DMNC | MNC | EPC |
|---|---|---|---|---|---|---|---|
| 1 | **EP300** | BRCA1 | BRCA1 | BRCA1 | BRCA1 | BRCA1 | BRCA1 |
| 2 | CREBBP | **EP300** | CREBBP | **EP300** | PML | PML | CREBBP |
| 3 | **TP53** | CTNNB1 | **EP300** | CREBBP | ABL1 | ABL1 | **EP300** |
| 4 | BRCA1 | **JUN** | DDX5 | ATM | BCL6 | BCL6 | DDX5 |
| 5 | CTNNB1 | **TP53** | SMARCA4 | SMARCA4 | ATM | ATM | ATM |
| 6 | SMARCA4 | CREBBP | ATM | RB1 | PTEN | PTEN | **TP53** |
| 7 | **JUN** | SMARCA4 | **TP53** | DDX5 | **EWSR1** | **EWSR1** | AKT1 |
| 8 | PTPN11 | AKT1 | AKT1 | AKT1 | AKT1 | AKT1 | SMARCA4 |
| 9 | **RB1** | RB1 | **JUN** | CTNNB1 | CREBBP | CREBBP | ABL1 |
| 10 | PIK3R1 | ABL1 | CTNNB1 | PIK3R1 | **EP300** | **EP300** | CTNNB1 |

**B)**

| Sr. no | Degree | Betweenness Centrality | Stress | MCC | DMNC | MNC | EPC |
|---|---|---|---|---|---|---|---|
| 1 | **TP53** | **TP53** | **TP53** | PTK2 | ERCC8 | **TP53** | **TP53** |
| 2 | PRKDC | YWHAZ | MAPK3 | PTK2B | FGFR1 | PRKDC | PRKDC |
| 3 | MAPK3 | PRKCA | PRKDC | MAPK3 | PDPK1 | MAPK3 | **MAPK8** |
| 4 | **MAPK14** | MAPK3 | PRKCA | PRKCA | IRS1 | **MAPK14** | MAPK3 |
| 5 | **MAPK8** | **MAPK14** | YWHAZ | SHC1 | RET | **MAPK8** | GRB2 |
| 6 | PRKCA | **PRKDC** | **MAPK14** | **GSK3B** | PTK2B | PRKCA | STAT3 |
| 7 | **GSK3B** | PCNA | PCNA | INSR | INSR | **GSK3B** | PRKCA |
| 8 | GRB2 | **MAPK8** | **MAPK8** | **MAPK14** | ERCC1 | GRB2 | **MAPK14** |
| 9 | YWHAZ | **GSK3B** | **GSK3B** | JAK2 | PDGFRB | YWHAZ | EGFR |
| 10 | **RB1** | PARP1 | GRB2 | **PRKDC** | HMGB2 | **RB1** | **GSK3B** |

**Table 3.** Overall statistics related to various subnets and SIRT interactome network

| Subnets | Average number of neighbors | Diameter | Average shortest path length | Network heterogeneity |
|---|---|---|---|---|
| **Sirt family network** | 84.12 | 5 | 2.6 | 1.385 |
| **Acetylation subnet** | 65.65 | 5 | 2.4 | 1.215 |
| **Aging subnet** | 25.3 | 4 | 2 | 0.726 |
| **Kinases subnet** | 23.35 | 4 | 1.9 | 0.913 |
| **Methylation subnet** | 15.16 | 4 | 2.1 | 0.891 |
| **Cancer subnet** | 11.11 | 5 | 2.7 | 1.019 |